\documentclass[pra,aps,twocolumn,groupedaddress,amsmath,amssymb,floatfix]{revtex4}
\usepackage{graphicx}

\newcommand{\be}{\begin{equation}}
\newcommand{\ee}{\end{equation}}
\newcommand{\xx}{{\mathbf x}}
\newcommand{\XX}{{\mathbf X}}

\newcommand{\kk}{{\mathbf k}}
\newcommand{\qq}{{\mathbf q}}
\newcommand{\QQ}{{\mathbf Q}}
\newcommand{\eq}[1]{(\ref{#1})}
\newcommand{\mume}{\mu \rm m}

\newcommand{\Tr}{{\rm Tr}}

\begin{document}
\title{Stochastic classical field model for polariton condensates}
\author{Michiel Wouters}
\affiliation{Insitute of Theoretical Physics, Ecole Polytechnique F\'ed\'erale de Lausanne (EPFL), CH-1015 Lausanne, Switzerland}
\author{Vincenzo Savona}
\affiliation{Insitute of Theoretical Physics, Ecole Polytechnique F\'ed\'erale de Lausanne (EPFL), CH-1015 Lausanne, Switzerland}
\begin{abstract}
We use the truncated Wigner approximation to derive stochastic classical field equations for the description of polariton condensates. Our equations are shown to reduce to the Boltzmann equation in the limit of low polariton density. Monte Carlo simulations are performed to analyze the momentum distribution and the first and second order coherence when the particle density is varied across the condensation threshold.
\end{abstract}
\maketitle

\section{Introduction}

 Condensates of microcavity polaritons~\cite{polarBEC_review} are a solid state realization of the two dimensional Bose gas. Their succesful creation relies on the peculiar nature of the microcavity polariton quasiparticle, that combine a very light effective mass (high quantum degeneracy temperature) with interparticle interactions that provide efficient relaxation. The formation of spontaneous coherence in these systems is now routinely achieved in several laboratories~\cite{yamamoto,kasprzak,snoke,krizhanovskii-prl08}.

One crucial difference between polariton condensates and other realizations of the two dimensional Bose gas such as liquid $^4$He films~\cite{bishop} and tightly confined ultracold atomic gases~\cite{dalibard} comes from the  finite life time of the microcavity polaritons of the order of a few ps. In order to compensate for the polariton losses, new particles can be continuously injected into the microcavity. The resulting steady state is not a thermal equilibrium one, still it shows features expected for an equilibrium BEC. For example, the tail of the momentum distribution can in many cases be fitted by an exponential Maxwell-Boltzmann decay. The lack of full thermalization is already clear from the fact that the extracted temperature is in general not equal to the temperature of the reservoir constituted by the semiconductor lattice~\cite{yamamoto,kasprzak,snoke}.

Effects that have no counterpart at equilibrium have been observed in polariton condensates. For example, the condensate state can depend dramatically on the size of the excitation spot~\cite{maxime,maxime-large}: in the case of a large pump spot the usual condensation around zero momentum is observed, instead for a small excitation spot the condensation occurs on a ring in momentum space. This difference has been explained within a mean field theory based on the Gross-Pitaevskii equation, including driving and dissipation~\cite{prb08}. 
More recently, another remarkable phenomenon related to the flow in a continuously pumped polariton condensate was observed experimentally~\cite{vortexnature}: vortices are spontaneously created in polariton condensates without setting the system into rotation. A theoretical interpretation of this effect was given in the framework of the  generalized Gross-Pitaevskii equation. In a significant fraction of random landscape realizations the polariton condensate contains a vortex.
A related prediction was made by Keeling and Berloff: they found that a rotating vortex lattice can be spontaneously generated in a large regular trap ~\cite{keeling-berloff}.

The above mentioned phenomena can be understood within a mean field theory, i.e. a theory where the quantum polariton field is replaced by a  classical field. In this approximation, all information on the fluctuations is however lost. Since we deal with a two-dimensional system, the physics of fluctuations in polariton condensates is in analogy with equilibrium systems expected to be very rich~\cite{thouless,dalibard,shlyapnikov2D} and the question arises for example to what extent the physics related to the Berezinskii-Kosterlitz-Thouless survives the driving and dissipation of polariton condensates.

\begin{figure}[htbp]
\begin{center}
\includegraphics[width=\columnwidth,angle=0,clip]{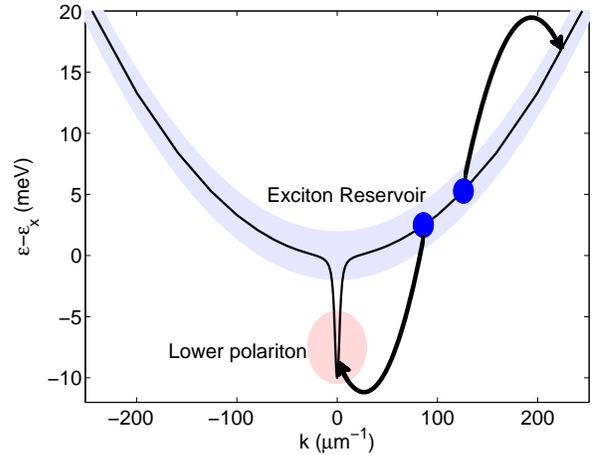}
\end{center}
\caption{Sketch of the separation of phase space in two regions: the lower polariton and the exciton reservoir. Scattering from the exciton reservoir into the lower polariton region replenishes the lower polaritons.}
\label{fig:sketch}
\end{figure}

Experimentally, the fluctuations of the polariton condensates have been investigated under the form of the first and second order coherence functions. In the first order equal time spatial  correlation function, long range correlations were observed above the stimulation threshold for condensation~\cite{kasprzak,yamamoto-g1}. Other correlation functions include the temporal first~\cite{krizhanovskii-prl08} and second order coherence ~\cite{krizhanovskii-prl08,yamamoto-g2,kasprzak-g2}.

The semiclassical Boltzmann equation~\cite{tassone,haug,malpuech} provides a theoretical description of the first order spatial coherence, which is the Fourier transform of the momentum distribution. Including the details of the relaxation mechanisms, this formalism is expeced to give a reliable estimate for the required polariton density to achieve condensation. Above the condensation threshold, the random phase approximation contained in the Boltzmann equation breaks down and more sophisticated techniques should be used. Schemes that have been implemented in the literature involve the separation of the condensate mode from the excited states~\cite{laussy} and a generalization of the Boltzmann equation that includes the coherences within a Bogoliubov approximation~\cite{sarchi}.

 One of the remarkable consequences of the nonequilibrium nature of the polariton condensates is that the collective excitation spectrum is changed at low wave vectors: a diffusive instead of soundlike character is found for the low energy phase modes. This dispersion of elementary excitations was found in a theoretical description based on a Keldysh Green function technique worked out by the Littlewood group~\cite{littlewood}. The same spectrum is straightforwardly recovered by linearizing the generalized Gross-Pitaevskii equation around a steady state~\cite{nonresonant}, a calculation that is easily extended to spatially nonuniform situations. 

It is well known from quantum optics and the theory of weakly interacting Bose gases that fluctuations can be included by introducing a stochastic element in the Gross-Pitaevskii equation~\cite{carlos,stoof-jltp,gardiner-zoller,zaremba}. For polariton systems in the parametric oscillation regime, such a method was used in Ref.~\cite{iac_threshold}.
One of the great advantages of these so called {\em classical field} methods is that the nonuniformity of a system does not introduce any appreciable extra cost in their numerical implementation. A second advantage is that practical numerical calculations do not require a perturbative expansion around a condensed state and can even be applied to study physics related to the condensation phase transition~\cite{iac-vort,simula-pra08}. Finally, these methods can describe the evolution of the system in real time so that information on both the steady state and transients can be obtained. The latter can be of particular use to model experiments that are performed under pulsed excitation~\cite{yamamoto}.

Due to the approximations involved in the classical field methods, they cannot describe the particles up to arbitrary large momenta, where quantum effects (most notably spontaneous scattering) are dominant~\cite{carlos}. In this respect, polariton condensates are very well suited for a classical field description, because, as illustrated in Fig.~\ref{fig:sketch}, the phase space can be naturally divided into two parts: i) a low energy polaritonic region with a small effective mass that shows quantum coherence above a certain threshold density and ii) a high energy excitonic `reservoir' region with a high effective mass that under typical experimental densities behaves as an incoherent classical gas. The role of the two subsystems is very different: the polaritonic field is the quantity of experimental interest because it is easily accessible in photoluminescence experiments and can be driven into the quantum degenerate regime. The role of the reservoir is to replenish the polariton region through relaxation.

We will present in this paper a set of classical field equations for the (Wigner distribution function of the) polariton dynamics coupled to the exciton reservoir and apply them to calculate the equal time first and second order coherence functions across the condensation threshold. For simplicity, we have not included the polarization degree of freedom. A Boltzmann description of polariton condensates including polarization can be found in Ref.~\cite{shelykh}. At thermal equilibrium, the magnitude of fluctuations can be parametrized by a single quantity, the temperature. For a weakly interacting system, this temperature can be extracted by fitting the tail of the momentum distribution with a Maxwellian curve.
Experiments have shown that the Maxwell-Boltzmann distribution is also recovered for the tail of the out of equilibrium polariton distribution~\cite{yamamoto,kasprzak,snoke} and has even been experimentally observed in the case of weak coupling lasing~\cite{blochlaser}. We will show that out of equilibrium, the universal characterization of fluctuations by a temperature parameter breaks down. We will point out several crucial aspects of the condensate-reservoir interactions that affect the correlation functions without changing the tail of the momentum distribution.

We start by presenting the model Hamiltonian for the nonresonantly excited polariton system in Sec. \ref{sec:ham}. It is shown in Sec. \ref{sec:master} how a master equation for the lower polariton field can be derived and how to solve it within the truncated Wigner approximation in Sec. \ref{sec:wigner}. The reservoir dynamics is discussed in Sec. \ref{sec:xres}. In Sec. \ref{sec:boltz}, we discuss the relation between our model and the Boltzmann equation. Numerical Monte Carlo results are presented in Sec. \ref{sec:numerics}. Conclusions are drawn in Sec. \ref{sec:concl}.

\section{Hamiltonian \label{sec:ham}}

In order to treat these two regions of polaritonic phase space with a very different character, we replace the original Hamiltonian by a Hamiltonian for polaritons and excitons, that are anihilated by the operators $\psi(\xx)$ and $\phi(\xx)$ respectively. In terms of these annihilation operators, our model Hamiltonian reads
\begin{equation}
	H = \int d \xx \left[ H_{LP}(\xx) + H_{R}(\xx) + H_{R,LP}(\xx)\right].
	\label{eq:Hsep}
\end{equation}
The lower polariton Hamiltonian density is the usual
\begin{equation}
	H_{LP}(\xx) = \psi^\dag(\xx) \frac{-\nabla^2}{2m_{LP}} \psi(\xx) 
	+ \frac{g}{2} \psi^\dag(\xx)\psi^\dag(\xx) \psi(\xx) \psi(\xx),
	\label{eq:HLP}
\end{equation}
where $m_{LP}$ is the lower polariton effective mass and $g$ quantifies the strength of the polariton-polariton interactions, that is well approximated by a zero-range potential. The exciton reservoir Hamiltonian is given by
\begin{equation}
	H_{R}(\xx) =  \phi^\dag(\xx) \frac{-\nabla^2}{2m_X}  \phi(\xx)
	+ \frac{g}{2}	\phi^\dag(\xx)  \phi^\dag(\xx) \phi(\xx) \phi(\xx). 
	\label{eq:hr}
\end{equation}
In the polariton/exciton basis, the exciton-exciton Coulomb scattering gives rise  to various coupling terms. The relevant ones are
\begin{equation}
		H_{R,LP}(\xx) = H_{R,LP}^{\rm loss}(\xx) + H_{R,LP}^{\rm gain}(\xx) + H_{R,LP}^{\rm mf}(\xx).
	\label{eq:hrlp}
\end{equation}
Lower polaritons are created by the term
\begin{equation}
	H_{R,LP}^{\rm gain}(\xx) = g\; \phi^\dag(\xx) \phi(\xx) \phi(\xx) \psi^\dag(\xx),
	\label{eq:hgain}
\end{equation}
whereas they are destroyed by
\begin{equation}
H_{R,LP}^{\rm loss}(\xx) = g\; \phi^\dag(\xx) \phi^\dag(\xx) \phi(\xx) \psi(\xx) 
	\label{eq:loss}.
\end{equation}

Mean field shifts of the lower polaritons due to the excitons in the reservoir and vice versa, are described by the Hamiltonian
\begin{equation}
	H_{R,LP}^{\rm mf}(\xx) = g\; \phi^\dag (\xx) \phi(\xx) \psi^\dag(\xx)\psi(\xx). 
	\label{eq:hmfs}
\end{equation}
Note that we have in fact extended phase space by introducing two particles: the excitonic phase space is extended down to $k=0$ and the polaritonic phase space to arbitrarily large momenta. Both extensions however add a very few states in the physically relevant regions.

\section{Master Equation \label{sec:master}}
In order to take advantage of the incoherent nature of the excitons in the reservoir, we will trace them out from the dynamics and obtain a quantum equation for the LP field alone.
The LP field dynamics can be studied through the Liouville equation for the density matrix 
\begin{equation}
	\dot{\rho} = -i [H, \rho].
	\label{eq:liouv}
\end{equation}
Going through the usual steps in the derivation of the Master equation in quantum optics, the Master equation for the full density matrix reads in the second Born approximation in $H_{R,LP}$
\begin{multline}
	\rho(t)  = \rho(t_0) - i \int_{t_0}^t dt' [H_{R,LP}(t'),\rho_{LP}(t')]\\
	- \int_{t_0}^t dt' \int_{t_0}^{t'} dt'' [H_{R,LP}(t'),[H_{R,LP}(t''),\rho(t'')]],
	\label{eq:mast0}
\end{multline}
where $H_{R,LP}(t)$ is in the interaction picture with respect to the Hamiltonian $H_0=H_{LP} + H_{R}$.

Taking the trace of this equation over the reservoir degrees of freedom gives the desired master equation for the reduced density matrix $\rho_{LP}$ of the lower polariton subsystem. The second term on the RHS vanishes when the trace over the reservoir is taken, so we only have to analyze the third one. It consists of terms like
\begin{equation}
\Tr_R\{H_{R,LP}^{\rm gain/loss} H_{R,LP}^{\rm gain/loss} \rho \} ,
\end{equation}
where 
\begin{equation}
\Tr_R\{H_{R,LP}^{\rm gain} H_{R,LP}^{\rm gain} \rho_{LP} \} =\Tr_R\{H_{R,LP}^{\rm loss} H_{R,LP}^{\rm loss} \rho_{LP} \} =0. 
\end{equation}
A nonzero term is e.g. given by
\begin{equation}
R_1=\int dt' \; dt'' \Tr_R\{H_{R,LP}^{\rm gain}(t') H_{R,LP}^{\rm loss}(t'') \; \rho(t'') \}. 
	\label{eq:1term}
\end{equation}
In order to work out the trace over the reservoir, we introduce relative and center of mass coordinates
\begin{eqnarray}
	\XX &=& \frac{\xx'+\xx''}{2}, \hspace{1cm}
	\xx = \xx'-\xx'',\\
	T &=& \frac{t'+t''}{2}, \hspace{1.2cm}
	t = t'-t''.
	\label{eq:relcom}
\end{eqnarray}
We define the Wigner transform of the reservoir propagator as
\begin{multline}
F^W(\XX,\kk,T,\omega)=\int d t \; d\xx \; e^{i\omega t} e^{-i \kk \xx} \\
	\Tr_R\{ \phi^\dag(\XX+\xx/2,T+t/2)\phi(\XX-\xx/2,T-t/2)\}.
	\label{eq:wignerF}
\end{multline}
With the inverse transformation, we obtain for $R_1$ defined in Eq.~\eq{eq:1term}
 \begin{multline}
	R_1=\frac{1}{\Omega^3}\sum_{\kk_{1,2,3}}\int d \XX\, d \xx  \; e^{i(\Delta \kk \cdot \xx-\Delta \epsilon t)} \; \Pi_f(\XX,\kk_{1,2,3}) \\
		\psi^\dag(\XX+\xx/2,T+t/2)  \psi(\XX-\xx/2,T-t/2) \rho(T-t/2),
	\label{eq:1termd}
\end{multline}
where $\Omega$ is the area of our system, $\Delta \kk=\kk_2+\kk_3-\kk_1$, 
$\Delta \epsilon =\epsilon(\kk_2)+\epsilon(\kk_3)-\epsilon(\kk_1)$ and $\Pi_f(\XX,\kk_{1,2,3})$ is a typical Boltzmann collision rate (density in phase space)
\begin{equation}
\Pi_f(\XX,\kk_{1,2,3}) =f(\XX,\kk_1,T)[f(\XX,\kk_2,T)+1][f(\XX,\kk_3,T)+1] 
\end{equation}
We have used the quasi-particle approximation \cite{haug_book}
\begin{equation}
	F^W(\XX,\kk,T,\omega) = (2\pi i) \delta(\omega-\epsilon_\kk) f(\XX,\kk,T).
	\label{eq:quasi-particle}
\end{equation}
The time evolution of the LP field operators is approximately given by
\begin{equation}
	\psi^\dag(\XX+\xx/2,T+t/2) \simeq \frac{1}{\Omega}\sum_\QQ e^{i\QQ(\XX+\xx/2)} e^{i\epsilon_{\QQ}t}
	\psi^\dag(\QQ,T-t/2), 
	\label{eq:psitimeshift}
\end{equation}
where the interaction shift in the frequency of $\psi_\QQ$ was neglected.
The exponential $e^{i\QQ \xx}$ can be combined with the exponential in Eq. \eq{eq:1termd}. Because the typical reservoir momentum is much larger than the typical lower polariton momentum (see Fig. \ref{fig:sketch}), this factor is negligible. For the same reason, also the $\xx$ in the second field operator in Eq. \eq{eq:1termd} can be neglected. If we then also assume that the density matrix is slowly varying on the microscopic time scale $t$, the integral over the relative time imposes energy conservation for the scattering process. We can then finally rewrite Eq. \eq{eq:1term} as
\begin{multline}
R_1=\pi g^2 T \int d \XX \frac{1}{\Omega^3}\sum_{\kk_1,\kk_2,\kk_3,\QQ} \delta_{\Delta\kk}
	\delta_{\Delta\epsilon_{\kk_1}+\epsilon_{\qq}} \\
	\Pi_f(\XX,\kk_{1,2,3})
	\psi_{\QQ}^\dag(T)  \psi_{\XX}(T) \rho(T) .
	\label{eq:1termc}
\end{multline}

The main simplifying assumption of the model consists now of assuming that the expression \eq{eq:1termc} is a function of the total reservoir density $n_R$ and the energy $\epsilon_\qq$ only. This comes down to the assumption of a steady state distribution of the reservoir excitons among the different $\kk$ states.

Working out the trace over the reservoir in Eq. \eq{eq:mast0} yields gain and loss terms for the lower polariton field from the collisons involving reservoir excitons. Collecting all these terms, we obtain
\begin{equation}
	\frac d{dt}\rho(t)  = -i[H_{LP},\rho] + K_{\rm in}(\rho) + K_{\rm out}(\rho), 
\label{eq:master1}
\end{equation}
where the density matrix evolves under the in-scattering as
\begin{multline}
K_{\rm in}(\rho)=  \frac{1}{2}\sum_{\qq}\int d\xx   R_{\rm in} (n_R,\epsilon_\qq)  
	\left[e^{i\qq \xx} \psi^\dag(\xx) \rho \psi(\qq)
\right. \\ \left.
- e^{i \qq \xx} \psi(\qq) \psi^\dag(\xx) \rho + {\rm h.c.} \right],
\end{multline}
and under out-scattering as
\begin{multline}
K_{\rm out}(\rho) = \frac{1}{2}\sum_{\qq}\int d\xx R_{\rm out} (n_R,\epsilon_\qq) \left[e^{i\qq \xx} \psi(\qq) \rho \psi^\dag(\xx)
\right. \\ \left.
 - e^{i\qq \xx} \psi^\dag(\xx) \psi(\qq) \rho + h.c. \right],
\end{multline}
The rates $R_{\rm in/out}$ are given by the usual semiclassical Boltzmann rates. Neglecting stimulated processes in the reservoir, $R_{\rm in}$ and $R_{\rm out}$ depend on the reservoir density respectively as $n_R^2$ and $n_R$. We therefore write 
\begin{eqnarray}
		R_{\rm in} (n_R,\epsilon_\qq) &=& n_R^2	R_{\rm in} (\epsilon_\qq)\\
		R_{\rm out} (n_R,\epsilon_\qq)&=& n_R R_{\rm out} (\epsilon_\qq)
	\label{eq:rnr}
\end{eqnarray}

Actually, another loss mechanism for the lower polariton field is present: leakage of the photon out of the imperfect microcavity mirrors, that gives a finite line width $\gamma$ to the lower polariton. This loss mechanism has a negligible energy and momentum dependence and can be added to the model by simply adding the constant term $\gamma$ to $R_{\rm out}$.

\section{Wigner \label{sec:wigner}}

An exact solution of the Master Equation \eq{eq:master1} is not possible, but numerical progress can be made by the use of quasi-probability distributions from quantum optics. In the presence of dissipation, the Wigner distribution function is believed to give robust results (see Ref. \cite{quant_noise}, pps. 115,124). This method has been applied to study BEC aspects of parametrically generated signal polariton in microcavities in Ref.~\cite{iac_threshold}.

The Wigner distribution function is a quasi probability distribution defined on the space spanned by the complex valued functions $\psi(\xx)$. In order to avoid ambiguity, we will use below an explicit `hat' notation for the quantum field operator $\hat \psi(\xx)$.

In terms of the density matrix, the Wigner distribution function is defined as
\begin{multline}
	P_W[\psi(\xx)] = \frac{1}{\pi^2}\int d^2 \lambda(\xx) 	
	\exp[\psi(\xx) \lambda(\xx)^* -\psi(\xx)^* \lambda(\xx)]\\
	\times \frac{1}{\pi} \int d^2 \alpha(\xx) 
	\langle \alpha(\xx) | 
	\rho \exp[\lambda(\xx) \hat \psi^\dag(\xx) -\lambda^*(\xx) \hat \psi(\xx) |\alpha(\xx) \rangle]  
	\label{eq:wignerdef},
\end{multline}
where $|\alpha(\xx) \rangle$ is a coherent state of polaritons at position $\xx$ with complex amplitude $\alpha(\xx)$. 
Expectation values calculated with the Wigner distribution function correspond to expectation values of symmetrized operator expressions. For example for the one-bbody density matrix, we have:
\begin{multline}
	\int d^2 \! \psi(\xx) \; P_W[\psi(\xx)]\; \psi^*(\xx)\psi(\xx') \\
	= \frac{1}{2} \Tr\{\rho [\hat \psi^\dag(\xx) \hat \psi(\xx')+\hat \psi(\xx')\hat \psi^\dag(\xx) \}.
	\label{eq:expectdens}
\end{multline}

Using the operator correspondences \cite{quant_noise}, the equation of motion for the Wigner quasi-probability distribution $P_W$ is computed:
\begin{widetext}
\begin{multline}
\frac{\partial P_W[\psi(\xx),\psi^*(\xx)]}{\partial t}
= \left\{ \frac{\partial }{\partial \psi(\xx)} F_{det} - \frac{\partial }{\partial \psi^*(\xx)} F^*_{det}
+\frac{\partial^2 }{\partial \psi(\xx)\psi^*(\xx)} 
[\gamma+\mathcal R_{\rm in}+\mathcal R_{\rm out }] \right. \\
\left. + i \frac{g}{2 \Delta V}
\frac{\partial^2 }{\partial \psi(\xx)\psi^*(\xx)}
\left[ \frac{\partial }{\partial \psi^*(\xx)} \psi^*(\xx)-\frac{\partial }{\partial \psi(\xx)}\psi(\xx)
 \right]
\right\}
P_W[\psi(\xx),\psi^*(\xx)].
\label{eq:wig}
\end{multline}
\end{widetext}
$F_{det}$ is the deterministic mean field force acting on the polaritons
\begin{equation}
F_{det}  =  -i \left[\frac{-\hbar^2 \nabla^2}{2m}+ \frac{i(\mathcal R_{\rm in}-\mathcal R_{\rm out}-\gamma)}{2}+
\frac{g}{\Delta V} |\psi(\xx)|^2 \right]\psi(\xx).
\end{equation}
In Eq.~\eq{eq:wig}, a momentum cutoff for the field $\psi$ is implicitely introduced by formulating the problem on a spatial grid with cell area $\Delta V$. The expression $\mathcal R_{\rm in,out} \psi$ should be understood as 
\begin{eqnarray}
	\mathcal R_{\rm in} \psi(\xx) &=&  n^2_R(\xx)\sum_{\qq} e^{-i\qq \xx'} R_{\rm in}(\epsilon_\qq)  \psi_{\xx'},\\
\mathcal R_{\rm in} \psi(\xx) &=& n_R(\xx)\sum_{\qq} e^{-i\qq \xx'} R_{\rm out}(\epsilon_\qq)  \psi_{\xx'}.
	\label{eq:kernel}
\end{eqnarray}
From the mathematical point of view, the last term in the equation of motion \eq{eq:wig} has proved to be an insurmountable problem. If this term is neglected (the so-called truncated Wigner approximation), the quasi-probability distribution $P_W$ obeys a standard Fokker-Plank equation, that correponds to the Langevin equation
\begin{equation}
d \psi(\xx) = F_{det} [\psi(\xx),\psi^*(\xx)]
+ dW(\xx),
\label{eq:stoch_mot}
\end{equation}
where $dW$ is a complex Gaussian stochastic variable with the correlation functions:
\begin{multline}
\langle dW(\xx) dW (\xx') \rangle =0, \hspace{1cm}\\
\langle dW(\xx) dW^* (\xx') \rangle =
 \frac{dt}{4 \Delta V} \left(\langle \xx| \mathcal R^S_{\rm  in}+\mathcal R^S_{\rm  out}| \xx' \rangle +2 \gamma \delta_{\xx,\xx'}\right),
\label{eq:noisecor}
\end{multline}
where $\mathcal R^S_{\rm in,out}=[\mathcal R_{\rm in,out}+(\mathcal R_{\rm in,out})^{\rm T}]/2$ are the symmetrized kernels. 

Let us now estimate the order of magnitude of the third order derivative in Eq. \eq{eq:wig} with respect to the other terms, in particular the second order derivative terms. The function $P_W$ is peaked around the value of the field $\psi(x)$ whose squared modulus equals $|\psi(\xx)|^2 = N(\xx)+1/2$. The variation of $P_W$ occurs on a scale of its argument of order one. Derivatives are therefore expected to be of the order of the function $P_W$ itself and the prefactors determine the relative importance of the derivative terms in Eq.(4). This leads us to the conclusion that the third order derivative is negligible with respect to the second order one if
\begin{equation}
\gamma \gg \frac{g}{\Delta V}.
\label{eq:wigcond}
\end{equation}
The dissipative character of the system thus increases the region of validity of the truncated Wigner approximation. The dissipation gives away information about the system and destroys nontrivial quantum states (e.g. number or Schr\"odinger cat states). In terms of the Wigner function, oscillations of $P_W$ accompanied by regions where it becomes negative (that cannot be represented by a regular probability distribution) are washed out by the dissipation~\cite{zurek}.

\section{The exciton reservoir \label{sec:xres}}

In our description of the microcavity dynamics, the exciton-like particles are treated as a classical reservoir. This approximation allowed to trace out the excitonic degrees of freedom and to isolate the quantum dynamics of the polaritons from the classical exciton dynamics. In principle, the reservoir density appears as a deterministic classical quantity in the resulting equations of motion for the lower polariton dynamics. Physically, however, this is not expected to be a very good approximation, because the condensate serves as a relaxation mechanism. Stimulated scattering makes the rate of this relaxation to depend on the reservoir population. 
Similar ideas have been implemented in Ref.~\cite{laussy}, where the dynamics of a single condensate mode was coupled to a Boltzmann equation for the excited states, and in Ref.~\cite{krizhanovskii-prl08}, where the reservoir was modelled by a saturable gain medium, a model widely used in laser physics~\cite{las-satgain}.

We propose to go beyond the approximation that the reservoir is unaltered by the system by coupling its dynamics to the equation of motion for the classical polariton field
\begin{equation}
\frac{d n_R}{dt}  = -\gamma_R [n_R - n^o_R(I_p,\psi) ],
\label{eq:motres}
\end{equation}
where $n^o_R(I_p,\psi)$ is the average steady state value of the reservoir density in the presence of a pump with intensity $I_p$ and a lower polariton field $\psi$. The relaxation time $\gamma^{-1}_R$ is a measure of the time it takes for the reservoir density to adjust to a new environment $(I_p,\psi)$. Spatial diffusion of the reservoir excitons is expected to be a small effect~\cite{nonresonant} and was therefore neglected.
For the steady state value of the reservoir density, we assume that it is simply proportional to the balance of incoming and outgoing particles
\begin{equation}
 n^o_R(P,\psi)= \beta (I_p-\frac{d}{dt} \langle \psi^\dag \psi \rangle \vert_{\rm res}),
\label{eq:nro}
\end{equation}
where $\frac{d}{dt}\langle \psi^\dag \psi \rangle \vert_{\rm res} = 2 {\rm Re}[\psi^* (\mathcal R_{\rm in}-\mathcal R_{\rm out})\psi]$ is the net scattering rate from the reservoir into the lower polariton branch.
It is instructive to substitute Eq. \eq{eq:nro} into Eq. \eq{eq:motres}:
\begin{equation}
 \frac{d n_R}{dt}  = P - \gamma_R n_R 
- \beta \gamma_R \frac{d}{dt} \langle \psi^\dag \psi \rangle \vert_{\rm res},
\end{equation}
where $P=\beta I_p$ is the effective pump term for the active reservoir polaritons. The parameter $\beta$  quantifies the backaction of the condensate on the reservoir. This backaction is needed to obtain a steady state for the dynamical equations above the threshold, where for $n_R=P/\gamma_R$ the in-scattering rate exceeds the out-scattering rate.
In mean field theory, the reservoir density $n_R$ is clamped to its threshold value $n_{R,mf}$ that satisfies for homogeneous systems 
$n_{R,mf}^2 R_{\rm in}(0)-n_{R,mf} R_{\rm out}(0) = \gamma$. 
If we rewrite the motion equations for $n_R$ in terms of the renormalized $\tilde n_R=n_R/n_{R,mf}$, we have 
\begin{equation}
 \frac{d \tilde n_R}{dt}  = \tilde P - \gamma_R \tilde n_R 
- \alpha \frac{d}{dt} \langle \psi^\dag \psi \rangle \vert _{\rm res},
\label{eq:motnorm}
\end{equation}
where $\alpha=\beta \gamma_R / n_{R,mf}$. Also in the presence of fluctuations, the dimensionless reservoir density $\tilde n_R$ is close to one above threshold, in order for the gain to compensate for the losses. The factor $\alpha$ plays an important physical role because the backaction of the condensate on the reservoir tends to damp the condensate fluctuations. If the condensate density is at some time larger than average, the reservoir will be depleted, $R_{\rm in}-R_{\rm out}$ decreases and the deterministic part in the equations of motion for the condensate will decrease the amplitude of the fluctuation. 
In principle, the parameter $\alpha$ could be calculated from the Boltzmann equation. We prefer however to study the physics in terms of this parameter, because it gives a good insight in the nonequilibrium aspects of the coherence.

In the truncated Wigner approximation, the density of polaritons is related to $\psi$ as $n=|\psi|^2-1/(2\Delta V)$, or in words, the classical field $\psi$ contains half a particle per mode of zero point fluctuations. These fluctuations should be taken into account when evaluating the last term in Eq.~\eq{eq:motnorm}. 
For the out-scattering, the zero-point fluctuations do not contribute and should be subtracted, whereas for the in-scattering, the zero-point fluctuations give rise to only half of the spontaneous in-scattering. The remaining part should be added.
The equation of motion for the reduced reservoir density then finally reads
\begin{multline}
 \frac{d \tilde n_R}{dt}  = \tilde P - \gamma_R \tilde n_R 
- \alpha \frac{d}{dt} \langle \psi^\dag \psi \rangle \vert _{res,W}\\
- \frac{\alpha}{2 \Delta V} \sum_\kk \left[R_{\rm out}(\epsilon_\kk)+ R_{\rm out}(\epsilon_\kk)\right].
\label{eq:motnormfin}
\end{multline}

\section{Relation with the Boltzmann equation \label{sec:boltz}}

In the dissipative case, the derivation of the truncated Wigner equation did not rely on the formation of a condensate. We can therefore describe within the same formalism the condensed and non-condensed polariton gas. In the case that coherence is negligible, it is instructive to simplify the stochastic equations of motion~\eq{eq:stoch_mot}. We will find that in the incoherent regime, the polariton condensate can be described with a Boltzmann-like equation~\cite{kagan}. 

For simplicity, we consider the case of a uniform reservoir density. By writing the stochastic motion equations for the field $\psi(\kk)$ in momentum space, treating the interactions in the second Born approximation, and assuming that there are no phase relations between the different momentum components  $\langle \psi^*(\kk,t)\psi(\kk',t) \rangle =[N (\kk,t)+1/2]\delta_{\kk,\kk'}$, one obtains the following Boltzmann like equation for the time evolution of the densities in momentum space
\begin{multline}
\frac{dN(\kk,t)}{dt} = R_{\rm in}(\epsilon_\kk) [N(\kk,t)+1]  -[R_{\rm out}(\epsilon_\kk)+\gamma] N(\kk,t) \\
+ I_B[N(\kk)] +I_Q[N(\kk)].
\label{eq:bmfin}
\end{multline}
The first two terms describe the evolution of the mode occupation due to the interaction with the reservoir and losses through the cavity mirrors. Collisions are described by the last two terms. $I_B$ is the usual Boltzmann collision integral:
\begin{multline}
I_B[N(\kk)] = -2\pi g^2 \sum_{\kk_1, \kk_2} \delta_{\gamma-R}(\epsilon_1+\epsilon_2-\epsilon_3-\epsilon_4)\\
\times\left[N_1N_2(1+N_3)(1+N_4)- N_3N_4(1 + N_1)(1 + N_2)\right],
\end{multline}
where $ N_1(t) =  N (\kk,t), N_2(t) = N (\kk_1 + \kk_2 - \kk,t), N_3(t) = N (\kk_1,t), N_4(t) = N(\kk_2,t)$ and analogous for the energies $\epsilon_i$. The $\delta$-function for energy conservation is broadened due to the finite lifetime of the polaritons:
$\delta_\nu(\omega) = \sin(\omega \nu)/(\pi \omega)$.

The extra collisional term is due to to the fact that our stochastic classical field model does not coincide with the true quantum dynamics (the third order derivatives in Eq. \eq{eq:wig} are neglected):
\begin{multline}
I_C [N(\kk)]= -\frac{\pi g^2}{2} \sum_{\kk_1, \kk_2} \delta_{\gamma-R}(\epsilon_1+\epsilon_2-\epsilon_3-\epsilon_4)\\
\times [N_1+N_2-N_3-N_4].
\end{multline}
This term is spurious, because the Boltzmann equation should be recovered in the incoherent limit. Our classical field model can therefore only be a good approximation of the full quantum dynamics if the term $I_C$ is negligible with respect to the other terms in Eq. \eq{eq:bmfin}. It scales as $I_C \propto (g/\Delta V)(g n \Delta V)$. If the occupation numbers per grid cell $n \Delta V$ are much larger than unity, the Boltzmann collision term $I_B$ is obviously dominant with respect to $I_C$. This is the typical condition for the use of the Wigner distribution function in the description of a stable Bose gas \cite{carlos}. 
For bosons with a finite life time, this condition can be relaxed, because even when $n \Delta V$ is not much larger than unity, the spurious term can be still much smaller than the reservoir term $R_{\rm in}$. The in-scattering rate $R_{\rm in}$ should compensate the losses $\gamma$. If occupation numbers are not large, the truncated Wigner is therefore still expected to yield physical results if $g/\Delta V \ll \gamma$. Note that the latter requirement coincides with the condition \eq{eq:wigcond} derived from the full equation of motion~\eq{eq:wig}.

If we neglect the collisional terms in Eq. (\ref{eq:bmfin}), the steady state solution is
\begin{equation}
 N(\kk) = \frac{1}{[\gamma+R_{\rm out}(\epsilon_\kk)]/R_{\rm in}(\epsilon_\kk)-1}.
\end{equation}
The simplest model that yields a temperature $T_R$ (that is in experiments typically higher than the lattice temperature) for the tail of the polariton momentum distribution is obtained by setting the out-scattering rate to zero
\begin{eqnarray}
R_{\rm out}(E) = 0,\\
R_{\rm in}(E) \propto \gamma \exp(-E/T_R).
\end{eqnarray}
Studies of the Boltzmann equation \cite{porras} have however shown that the rates $R_{\rm in}$ and $R_{\rm out}$ both tend to increasse as a function of the energy, approximately as
\begin{eqnarray}
R_{\rm in} \propto \exp(E/k_B T_R),\\
R_{\rm out} \propto \exp(2 E/k_B T_R).
\label{eq:ratesporras} 
\end{eqnarray}
We will see below that a nonzero out-scattering enhances fluctuations.

\section{Numerical Results \label{sec:numerics}}

The stochastic motion equations can be simulated by Monte Carlo techniques. As we have already mentioned in the introduction, the nonequilibrium condition of the polariton condensates makes that the effect of the reservoir on the correlation functions cannot be characterized by the temperature alone. We will discuss below two other physical quantities that determine the degree of coherence in the polariton condensate: the feedback parameter $\alpha$ and the out-scattering rate $R_{\rm out}$. The other parameters, we keep fixed for all simulations: $m/\hbar = 1 \mume^{-2} {\rm meV}^{-1}$, $g/\hbar=0.03 \mume^{2}$ and $k_B T_R=2 {\rm meV}$. The simulations were done on a $32\times 32$ point grid with physical dimension of $66\times 66 \mume^2$ and periodic boundary conditions.

\begin{figure}[htbp]
\begin{center}
\includegraphics[width=\columnwidth,angle=0,clip]{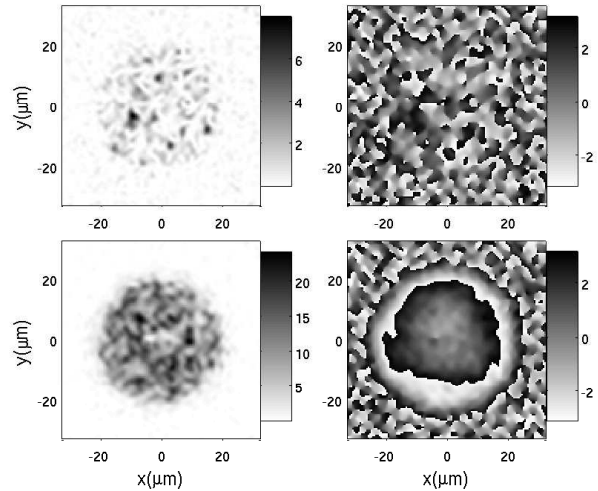}
\end{center}
\caption{Snapshots of a single Monte Carlo realization of the density (upper panels) and phase (lower panels) for a finite size excitation spot with intensity below (left hand panels) and above threshold (right hand panels).}
\label{fig:nphifin} 
\end{figure}

\begin{figure}[htbp]
\begin{center}
\includegraphics[width=\columnwidth,angle=0,clip]{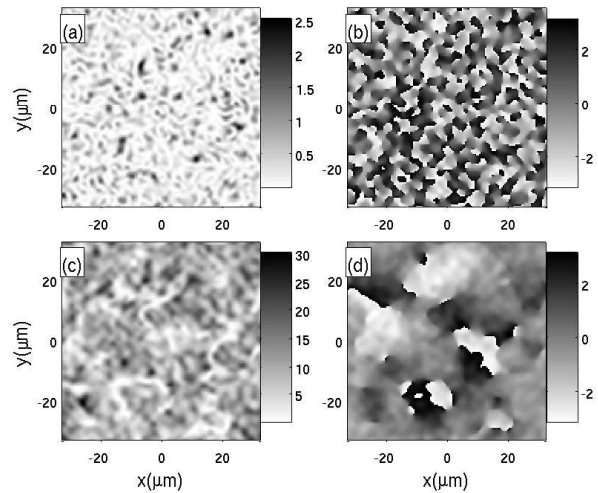}
\end{center}
\caption{Snapshots of a single Monte Carlo realization of the density (a,c) and phase (b,d) for excitation parameters below (a,b) and above threshold (c,d).}
\label{fig:nphipbc} 
\end{figure}

Figs. \ref{fig:nphifin} and \ref{fig:nphipbc} illustrate single Monte-Carlo realizations of the classical field $\psi(\xx)$. Even though these images have strictly speaking no direct physical meaning, they already illustrate qualitatively the coherence properties of the polariton condensate.
Fig. \ref{fig:nphifin} shows two examples for a finite excitation spot for pump intensities below (panels a and b) and above the threshold (panels c and d). At low density, both the density and phase fluctuations are large, whereas the phase fluctuations are clearly suppressed in the high density regime. 
Panel (d) shows that phase coherence exists all over the extent of the excitation region. The concentric phase profile originates from the repulsive polariton-polariton interaction that causes an outward flow of polaritons~\cite{prb08}.

Figure~\ref{fig:nphipbc} shows snap shots of the polariton density and phase for a uniform pump below and slightly above the threshold. The phase profile of panel (d) shows that the phase ordering is only partial. Vortex-anti vortex pairs appear to exist at densities well above the stimulation threshold. This is an indication that the physics of the Berezinskii-Kosterlitz-Thouless type could occur in polariton condensates.

\begin{figure}[htbp]
\begin{center}
\includegraphics[width=0.6\columnwidth,angle=0,clip]{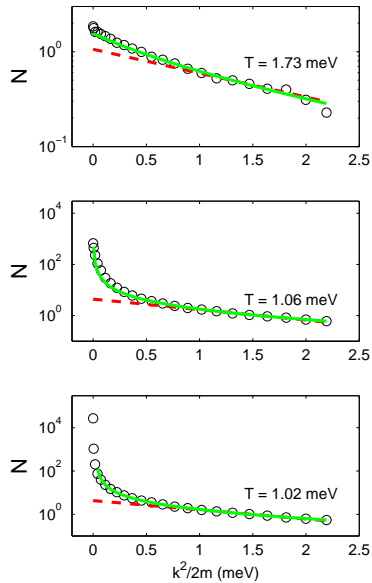}
\end{center}
\caption{Momentum distribution of the polariton field for various pump intensities. The open circles are the result of the Monte Carlo simulations and the full line is a fit to a BE distribution. Statistical errors of the Monte Carlo simulations are within the symbol size.}
\label{fig:nk} 
\end{figure}

Three momentum distributions for increasing pump intensity are shown in Fig. \ref{fig:nk}. As expected, our model shows the build up of a large occupation of the low momentum states for increasing pump intensity. The momentum distributions appear to be rather well fitted by a Bose-Einstein function (full line). It is important to mention here the important role of the reservoir relaxation rates. We have chosen them in such a way that a thermal distribution is obtained even in the absence of collisions between  lower polaritons. In simulations with energy independent relaxation rates (not shown) and a large, yet realistic ~\footnote{an experimental upper bound to the blue shift due to polariton-polariton interaction is given by the total blue shift, which is less than 1 meV} polariton-polariton interaction strenth, we have obtained a constant instead of exponential decay at large momenta.

Note that the temperature extracted from the fits of the tails to a Maxwellian is lower than the reservoir temperature $T_R$ (2 meV for the present simulations), that enters the rates $R_{\rm in,out}$ according to Eq. \eq{eq:ratesporras}: the nonlinearity modifies the temperature that is expected in the linear regime. We remind the reader that $T_R$ does not coincide with the lattice temperature and that $T_{\rm fit}<T_R$ does not imply that the polariton temperature is lower than the lattice temperature.

\begin{figure}[htbp]
\begin{center}
\includegraphics[width=1\columnwidth,angle=0,clip]{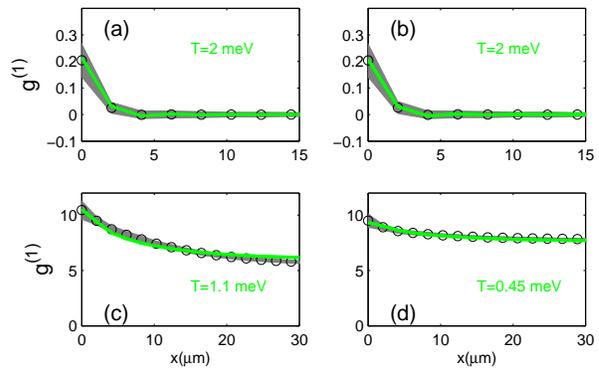}\\
\end{center}
\caption{Spatial coherence (open circles, the grey band indicates the error on the Monte Carlo data) for two values of the feedback from the condensate on the reservoir: from left to right $\alpha=0.01\mume^2$ (a,c) and $\alpha=0.1 \mume^2$ (b,d) . The full line shows the decay of the correlations in a non-interacting Bose gas, at the temperature reported in the panels.}
\label{fig:g1_2x} 
\end{figure}

The subtle features of long range coherence are much clearer in the Fourier transform of the momentum distribution, i.e. the first order spatial coherence function. In Fig.~\ref{fig:g1_2x} two values of $\alpha$ are compared. Below the condensation threshold, the gain saturation parameter $\alpha$ (see Eq. \eq{eq:motnormfin}) has no influence and the fit of the coherence by the $g^{(1)}$ of the noninteracting Bose gas yields the reservoir temperature of 2 meV.
For the simulation above the threshold, a higher value of $\alpha$ improves the long range coherence. Both spatial coherence functions are relatively well fitted by the one of a noninteracting Bose gas. Both temperatures are below the reservoir temperature. The lowest effective temperature is obtained for the largest feedback parameter $\alpha$.

\begin{figure}[htbp]
\begin{center}
\includegraphics[width=1\columnwidth,angle=0,clip]{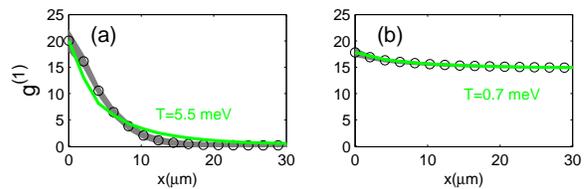}\\
\end{center}
\caption{The spatial coherence function as in Fig.~\ref{fig:g1_2x}, but including out-scattering. Simulatios with $\alpha=0.01 \mume^{2}$ (a) and  $\alpha=0.1 \mume^{2}$ (b) are shown.}
\label{fig:g1_Rout} 
\end{figure}

In the simulations of Fig.~\ref{fig:g1_2x}, the out-scattering was set to zero. In the simulations presented in Fig.~\ref{fig:g1_Rout}, we have included this effect. In order to avoid exceedingly large rates in the model, we have put a cutoff in the magnitude of 
$R_{\rm out}$ as $R_{\rm out}(E)={\rm min}[e^{2 E/ T_R},3.3 {\rm meV}]$. The in-scattering rate was chosen $R_{\rm in}(E)=[R_{\rm out}(E)+\gamma]e^{- E/ T_R}$.

Fig.~\ref{fig:g1_Rout} shows that the out-scattering has a big effect on the coherence function. This should not come as a surprise, because the out-scattering increases the fluctuations (physically shot noise due to the discrete nature of the polariton field). Keeping $\alpha=0.01 \mume^{-1}$ as in Fig.~\ref{fig:g1_2x}, but including some out-scattering, the coherence in panel (a) is dramatically decreased. As compared to the simulations of Fig.~\ref{fig:g1_2x}, the effect of $\alpha$ is much more pronounced. For the smallest value of $\alpha$, the temperature of 5.5 meV for which the spatial coherence is reasonably well fitted, is much larger than the one that is extracted from the tail of the momentum distribution (less than 1 meV): the polariton condensate behaves in this regime very different from the ideal Bose gas.

Another quantity of great physical interest is the second order coherence function $g^{(2)}(\xx,t;\xx',t')=\langle \psi^\dag(\xx,t) \psi^\dag(\xx',t') \psi(\xx',t') \psi(\xx,t) \rangle$, that quantifies the density fluctuations. Experimentally, the equal position second order coherence was investigated, in Ref.~\cite{kasprzak-g2} for equal times $t=t'$ and in Ref.~\cite{krizhanovskii-prl08} as a function of the delay $t-t'$. Within the Wigner formalism, different time correlation functions are not straighforwardly calculable, so we present here only results for the equal time second order coherence.

\begin{figure}[htbp]
\begin{center}
\includegraphics[width=\columnwidth,angle=0,clip]{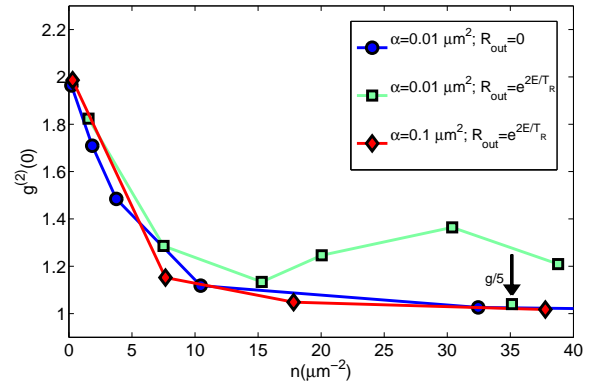}
\end{center}
\caption{Second order coherence function as a function of the total density obtained for different excitation powers and the simulation parameters indicated in the legend.}
\label{fig:g2} 
\end{figure}

Results of the equal position second order coherence $g^{(2)}(0)\equiv g^{(2)}(\xx,t;\xx,t)$ are shown in Fig.~\ref{fig:g2}, for several parameter values. As expected, $g^{(2)}(0)$ approaches the value 2 of the incoherent Bose gas in the low density regime. For increasing polariton densities, the second order coherence decreases, but its actual value depends again strongly on the chosen parameter values. 
A larger value of the feedback parameter $\alpha$, suppresses the density fluctuations. This is in agreement with the model described in Ref.~\cite{krizhanovskii-prl08}, where the density fluctuations are proportional to the saturation density (large saturation density means small feedback from the condensate on the reservoir).
Fig.~\ref{fig:g2} also shows that the out-scattering increases the density fluctuations. This dependence is expected, because adding the knock out processes leaves the deterministic term in the evolution equation for the classical field unaltered, but increases the fluctuations.

Note that the density fluctuations are within our model not always monotonous, but for some parameter values show a minimum value slightly above threshold. Nonmonotonous behavior of $g^{(2)}(0)$ was also observed in experiments on polariton condensation in CdTe microcavities~\cite{kasprzak-g2}. Also in the theoretical work of Ref.~\cite{paolo-g2,davide-g2} based on a Boltzmann equation for the excited states coupled to a master equation for the condensate mode, an increase of  density fluctuations above the threshold was found. It is however important to mention that in Fig. \ref{fig:g2} the interaction energy is very large when $g^{(2)}(0)$ increases again (1 meV blue shift due to condensate-condensate interactions alone). When the value of the blue shift is reduced to 0.2 meV, $g^{(2)}(0)$ is found to be very close to one.

We want to point out that we have not found any regime with good long range spatial coherence and large density fluctuations. Indeed, Fig.~\ref{fig:g1_Rout} (a) shows that at the density $n\approx 20 \mume^{-2}$ the spatial coherence is, although longer than the thermal de Broglie wave lenth corresponding to $T_R$, limited to about 10 $\mume$. Physically it is actually not expected that good spatial coherence and strong density fluctuations can go together, because phase fluctuations are coupled to the density fluctuations through the interaction and kinetic energy. So far, in the experiments on CdTe microcavities where the increase of $g^{(2)}(0)$ as a function of pump power was observed, no decrease in spatial coherence was seen. It is possible that the distance at which the spatial coherence was probed is too short for the decrease in spatial coherence to be detectable, but we cannot exclude other explanations in terms of extrinsic experimental effects. The measured density fluctuations could for example contain a component due to intensity fluctuations in the excitation laser.

\section{Conclusions \label{sec:concl}}

We have derived classical field equations for a nonresonantly excited polariton condensate in a semiconductor microcavity within the truncated Wigner approximation. Thanks to the polariton losses our model remains physical in the low density regime and allows to describe the polariton condensate at all densities.
Our equations were shown to reduce to the Boltzmann equation in the low density regime below threshold. Above threshold, the equations were analyzed numerically with Monte Carlo simulations. The first and second order spatial coherence were shown to depend dramatically on the feedback from the condensate on the reservoir (the gain saturation) and on the collisions with reservoir excitons that knock polaritons out of the condensate.
Within our model, the density fluctuations can show nonmonotonous behavior as a function of the polariton density. We predict that an increase in density fluctuations is accompanied by a decrease in the spatial coherence.

Finally, the vortex defects in individual Monte Carlo realizations of the polariton field show that the spatial coherence is limited by the spontaneous appearance of vortex defects in the phase. A further study of the role of vortices is necessary to understand their effect on the spatial coherence.

\section{Acknowledgements}

It is a pleasure to acknowledge numerous stimulating and insightful discussions with I. Carusotto, D. Sarchi, K. Lagoudakis, M. Richard and B. Pietka.

\end{document}